\documentclass[prb,aps,reprint,floatfix,superscriptaddress]{revtex4-1}
\usepackage{graphicx}
\usepackage{subfigure}
\usepackage{natbib}
\usepackage{makecell}
\usepackage{multirow}

\begin{document}

\title{Bonding Charge Density and Ultimate Strength of Monolayer Transition Metal Dichalcogenides} 

\author{Junwen Li}
\affiliation{Department of Materials Science and Engineering, University of Pennsylvania, Philadelphia, PA 19104, USA}
\author{Nikhil V. Medhekar}
\affiliation{Department of Materials Engineering, Monash University, Clayton, Victoria 3800, Australia}
\author{Vivek B. Shenoy}
\email{vshenoy@seas.upenn.edu}
\affiliation{Department of Materials Science and Engineering, University of Pennsylvania, Philadelphia, PA 19104, USA}
\date{\today}

\begin{abstract}
Two-dimensional~(2D) semiconducting transition metal dichalcogenides~(TMDs) can withstand a large deformation without fracture or inelastic relaxation, making them attractive for application in novel strain-engineered and flexible electronic and optoelectronic devices. In this study, we characterize the mechanical response of monolayer group VI TMDs to large elastic deformation using first-principles density functional theory calculations. We find that the ultimate strength and the overall stress response of these 2D materials is strongly influenced by their chemical composition and loading direction. We demonstrate that differences in the observed mechanical behavior can be attributed to the spatial redistribution of the occupied hybridized electronic states in the region between the transition metal atom and the chalcogens. In spite of the strong covalent bonding between the transition metal and the chalcogens, we find that a simple linear relationship can be established to describe the dependence of the mechanical strength on the charge transfer from the transition metal atom to the chalcogens.
\end{abstract}

\pacs{}

\maketitle

\section{INTRODUCTION}
Since its first mechanical exfoliation from graphite, graphene --- an allotrope of carbon with a two-dimensional~(2D) structure --- has attracted tremendous scientific and technological attention due to novel electronic, mechanical, and chemical properties.\cite{Novoselov_science_2004, Novoselov_pnas_2005} The intense interest in graphene has also stimulated an active search for possible inorganic 2D or quasi-2D materials with unique characteristics.  Examples of such materials include hexagonal boron nitride, transition metal dichalcogenides~(TMDs), and transition metal oxides.\cite{Lin_nanoscale_2012, Wang_naturenano_2012, Balendhran_advm_2013} Earlier attempts to obtain monolayer and a few-layer thick samples of these materials were based on mechanical exfoliation from their bulk counterparts. Recent advances in fabrication techniques, such as liquid exfoliation, lithium intercalation, epitaxy, laser thinning, and chemical vapor deposition, have successfully resulted in a scalable production of large 2D samples containing only a single layer to a few layers. \cite{Coleman_science_2011, Joensen_mrb_1986,Schumacher_ss_1993, Shi_nl_2012, Castellanos_nl_2012,Mattevi_jmc_2011}

Among the non-carbon 2D materials that are being explored, dichalcogenides of group IV, V and VI transition metals are of particular interest. Similar to graphene and its bulk counterpart graphite, these materials have strong in-plane bonding while the individual layers are bonded by weak van der Waals interactions. Several TMDs, unlike graphene, have an intrinsic band gap in the range of 1 - 2 eV, making them attractive for field effect transistors and optoelectronic devices.\cite{Wilson_aip_1969, Radisavljevic_nnano_2011} Many single-layer TMDs also demonstrate novel electronic properties that are otherwise not seen in their bulk or few-layer thick counterparts, for example, direct to indirect band gap transition, enhanced photoluminescence, and valley polarization.\cite{Splendiani_nl_2010, Eda_nl_2011, Ellis_apl_2011, Zhu_prb_2011,Cao_naturecomm_2012,Xiao_prl_2012,Zeng_naturenano_2012,Mak_naturenano_2012}

Recent mechanical experiments show that 2D TMDs can sustain very large elastic strains~($\sim$ 10\% effective in-plane strain) with a high resistance to inelastic relaxation and fracture, and hence these materials are also of interest for flexible electronics applications.\cite{Bertolazzi_acsnano_2011} Since the electronic conduction and valence energy states both depend on the strain, it has also been proposed that elastic strain can be employed to tune the band gap in monolayer TMDs.\cite{Yue_pla_2012, Yun_prb_2012,Priya_acsnano_2012} In order to realize the potential applications of various 2D TMDs in strain-engineered and flexible devices, an accurate characterization of their ability to withstand large elastic deformation is crucial.

Recently, Bertolazzi \textit{et al.}\cite{Bertolazzi_acsnano_2011} performed the experimental measurements on the stiffness and breaking strength of monolayer MoS$_2$ by using an atomic force microscope tip to deform the monolayer MoS$_2$ placed on a prepatterned SiO$_2$ substrate. They found that the effective Young's modulus and average breaking strength of monolayer MoS$_2$ are $270 \pm 100 \; \mbox{GPa}$ and 23~GPa, respectively. Using a similar technique, Castellanos-Gomez \textit{et al.}\cite{CGomez_am_2012} studied the mechanical properties of freely suspended MoS$_2$ nanosheets with 5 to 25 layers and obtained the mean Young's modulus of $330 \pm 70$~GPa.  These experiments clearly establish 2D MoS$_2$ as an ultra-strength material able to withstand large elastic deformation. However, such studies provide only effective orientation-averaged values of the mechanical properties and are limited only to single composition, while the rest of the 2D semiconducting TMDs remain yet to be characterized. The theoretical studies of mechanical properties of monolayer TMDs are also limited to the calculation of simple elastic properties that are relevant only at small deformations.\cite{Yue_pla_2012,Ataca_jpcc_2011} What is the intrinsic ultimate strength --- the maximum stress a defect-free material can withstand without failure --- of monolayer TMDs along different crystal directions and how is it dependent on the chemical composition? Indeed, the relationship between the intrinsic stress response of monolayer TMDs to large elastic deformation, the atomic-level geometry, and the chemical composition remains unclear.

In this article, using first-principles density functional theory simulations,  we report on the fundamental mechanical behavior of monolayer TMDs when subjected to a large deformation. Here we consider TMDs with chemical composition  MX$_2$, where M~(= Mo, W) is a group VI transition metal and X~(= S, Se, Te) is a chalcogen. Among the range of naturally occurring TMDs, group VI TMDS are semiconductors and, therefore, are good candidates for flexible and strain-engineered optoelectronic and photonic devices. Here we have determined the entire stress response of monolayer TMDs to large applied elastic strains --- the intrinsic materials characteristics that are crucial in designing the strain-engineered and flexible devices. We find that the stress response and the ultimate strength of monolayer TMDs strongly depends on the chemical composition as well as the loading direction. For the same transition metal M, the sulfides~(MS$_2$) are the strongest and selenides (MSe$_2$) are stronger than the tellurides (MTe$_2$). The chalcogens of W (WX$_2$) can accommodate larger stresses than that of Mo (MoX$_2$). For all compositions, the ultimate strength of the monolayer along the armchair direction far exceeds the strength in zigzag direction, typically by a factor of 1.5 - 2.2.  Our electronic structure analysis reveals that the origin of the observed mechanical behavior can be attributed to the hybridization between M-$d$ and X-$p$ orbitals and the charge transfer from the transition metal to the chalcogens.

\section{RESULTS AND DISCUSSION}

\begin{figure}  
  \includegraphics[scale=0.6]{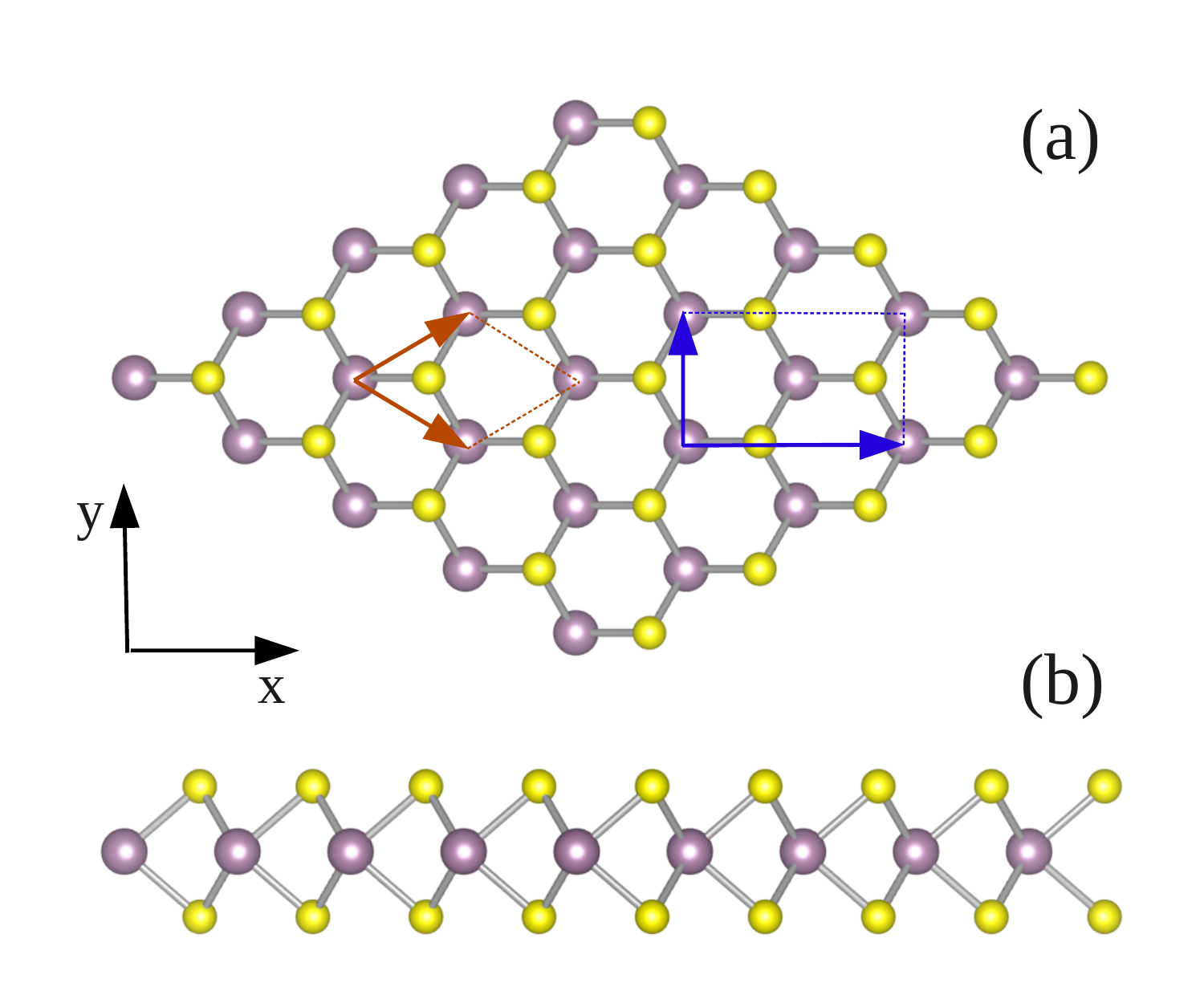}
  \caption{A schematic showing the crystal structure of a TMD monolayer~(MX$_2$) with (a) top view and (b) side view. The transition metal~(M) and chalcogen~(X) atoms are represented by purple (large) and yellow (small) spheres, respectively. The unit cell~(red) and the orthogonal supercell~(blue) are also depicted in (a). The armchair and zigzag orientations correspond to $x$ and $y$ directions, respectively.}
  \label{fig:scell}        
\end{figure}

All MX$_2$ TMDs monolayers considered in our study share a similar hexagonal crystal structure as shown in Fig.~\ref{fig:scell}.  A single layer of MX$_2$ consists of three X-M-X sublayers, with hexagonal lattice plane of the transition metal M sandwiched between identical hexagonal planes of the chalcogens X. In order to evaluate the mechanical response of the pristine TMD monolayers, we carried out first-principles density functional calculations within the local density approximation as implemented in ABINIT.\cite{Gonze_cpc_2009, Gonze_zk_2005}  Norm-conserving pseudopotentials in the form of Hartwigsen-Geodecker-Hutter pseudopotentials\cite{Hartwigsen_prb_1998} including the semicore electrons as valence electrons for Mo and W were used to describe the interaction between core and valence electrons. An energy cutoff of 65 Hartree was used for plane wave basis expansion. The $k$-point sampling schemes of $15 \times 15 \times 1$ and $10 \times 15 \times 1$ were employed during the geometry optimization of strain-free monolayer sheets and  the calculation of stress-strain relations, respectively.

For each of the MX$_2$ monolayers, we first performed the structure relaxation including the lattice constants and atomic coordinates as summarized in Table~\ref{tab:lattice}.  The lattice constants obtained from our DFT calculations are in excellent agreement with the experimental measurements~(error less than 1\%).\cite{Wilson_aip_1969} Our results are also in close agreement with those computed with different methods such as numerical atomic orbitals,\cite{Kumar_epjb_2012} PAW potentials\cite{Ding_pbcm_2011}. 

\begin{table}
  \caption{Lattice constants, M-X bond lengths and X-X distances for monolayer MX$_2$~(M = Mo, W; X = Se, Se, Te) TMDs.}
  \begin{ruledtabular}
  \begin{tabular}{cccc}
    MX$_2$ & $a$~(\AA) &   M-X~(\AA)   & X-X~(\AA)\\
    \hline
    MoS$_2$ & 3.13 & 2.38 & 3.11  \\
    MoSe$_2$ & 3.25 & 2.51 & 3.32  \\
    MoTe$_2$ & 3.49 & 2.70 & 3.60 \\
    
    WS$_2$ & 3.13 & 2.39 & 3.12 \\
    WSe$_2$ & 3.25 & 2.51 & 3.33  \\
    WTe$_2$ & 3.49 & 2.70 & 3.61     
  \end{tabular}
  \end{ruledtabular}
  \label{tab:lattice}
\end{table}

\begin{figure*}[t]
  \includegraphics[scale=0.35]{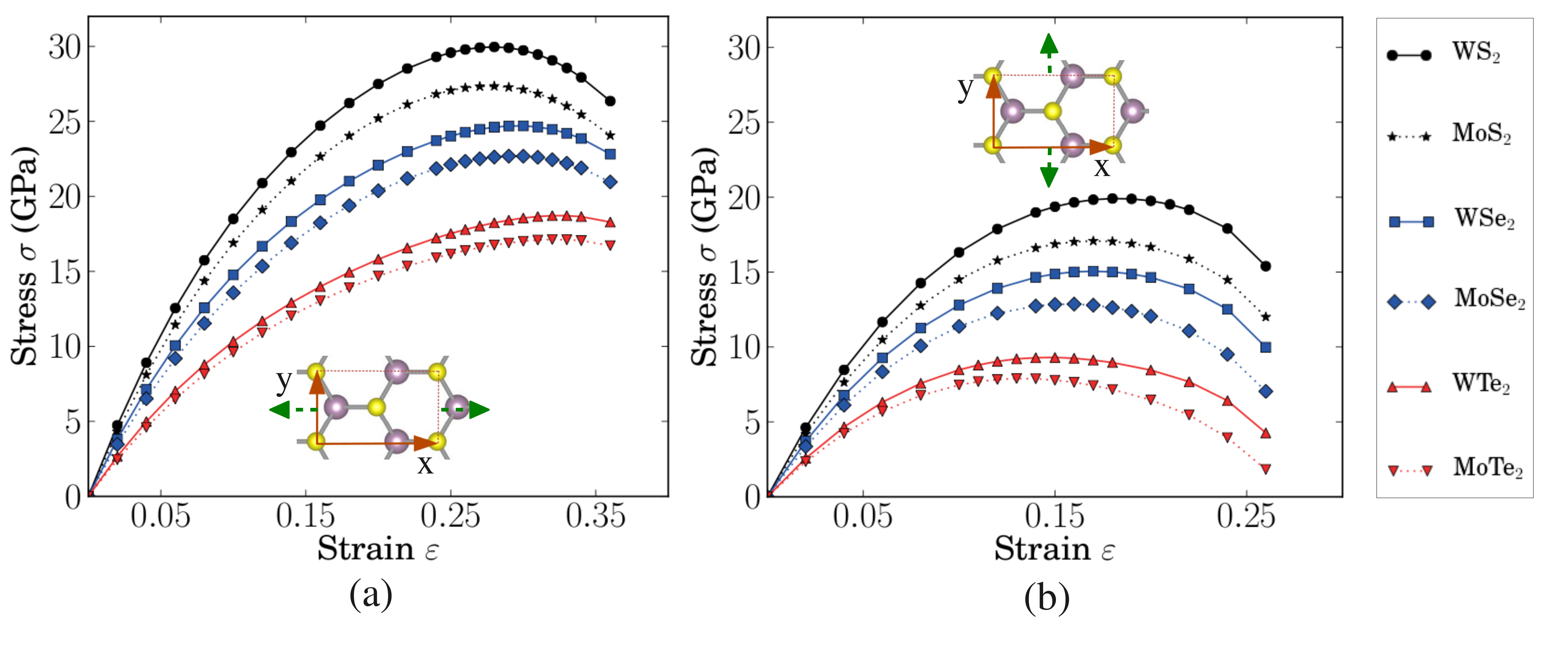}
  \caption{Tensile stress $\sigma$ as a function of uniaxial strain $\varepsilon$ along (a) armchair and (b) zigzag directions, respectively, for monolayer MX$_2$~(M = Mo, W; X = S, Se, Te) TMDs. Solid and dashed lines are used for WX$_2$ and MoX$_2$, respectively.}
  \label{fig:stress}
\end{figure*}

Similar to the 2D hexagonal lattice of graphene, two high-symmetry directions can be identified in the crystal structure of monolayer TMDs, namely, armchair and zigzag directions, which are oriented along and perpendicular to M-X bonds when projected on the plane containing transition metal atoms, respectively. Since the mechanical behavior of a defect-free material is ultimately controlled by the strength of its chemical bonds, it can be expected that the stress response of the monolayers along the armchair and zigzag directions will also be distinct. We therefore employed an orthogonal supercell as shown in Fig.~\ref{fig:scell}(a) to calculate the stress-strain curves along the two high symmetry directions. In each case, the super-cell was deformed along the high symmetry direction in small strain increments and the in-plane supercell vector normal to the the applied strain was allowed to relax in order to account for the Poisson's contraction as typically observed in the experiments.
The supercell stress as output directly by ABINIT is the stress averaged over the supercell volume. We renormalized the calculated supercell stress by a factor of $Z/h$, where $Z$ is the vacuum distance 12~\AA\ used to avoid the interaction between periodic images and $h$ is the interlayer distance in the bulk MX$_2$. The strain is defined as $\varepsilon = \frac{L-L_0}{L_0}$, where $L_0$ and $L$ are the supercell length along armchair or zigzag directions before and after applying tensile strain, respectively.

In Figs.~\ref{fig:stress}(a) and~\ref{fig:stress}(b) we present the calculated stress-strain relations along armchair~($x$) and zigzag~($y$) directions, respectively. It can be readily noted that for small strain, the stress for all MX$_2$ exhibits linear dependence on the applied strain for both loading directions. Table~\ref{tab:xx-yy} presents the Young's modulus $E$ calculated as the slope of the stress-strain curve in the region of strain less than 4\%. It is evident that for all compositions, the values of the Young's modulus obtained independently from the stress-strain curve for armchair and zigzag loading direction are virtually identical. This behavior is consistent with the symmetry of the MX$_2$ crystal lattice - the Young's modulus and other second order elastic constants are essentially isotropic due to the hexagonal symmetry in the basal plane. Our calculated values for the Young's modulus are also in good agreement with earlier studies. For instance, we obtained the Young's modulus of MoS$_2$ to be $\sim 220$~GPa, while experimental measurements have reported values of $270 \pm 100 \; \mbox{GPa}$  and $330 \pm 70 \; \mbox{GPa}$ for monolayer ~\cite{Bertolazzi_acsnano_2011} and few layer MoS$_2$ sheets~\cite{CGomez_am_2012}, respectively. 

\begin{table*}
  \caption{Calculated Young's modulus $E$~(GPa), ultimate strength $\sigma^*$~(GPa), ultimate strain $\varepsilon^*$ corresponding to the ultimate strength for armchair and zigzag directions, anisotropy factor $\phi$ and the charge transfer $\Delta Q$~($e$) from M to X for strain-free MX$_2$ monolayer sheets. }
  \begin{ruledtabular}
  \begin{tabular}{ccccccccc}
    \multirow{2}*{MX$_2$} & \multicolumn{3}{c}{Armchair~($x$)} & \multicolumn{3}{c}{Zigzag~($y$)} & \multirow{2}*{$\phi$} & \multirow{2}*{$\Delta Q$}\\ [-0.0ex]    
    &  $E$ & $\sigma^*$  & $\varepsilon^*$ & $E$ & $\sigma^*$  & $\varepsilon^*$ &    \\\hline 
    MoS$_2$ &   222.75  & 27.35    & 0.28     &  219.46  & 16.90    & 0.19     & 1.62 & 0.92 \\
    MoSe$_2$ &  178.78  & 22.68    & 0.29     &  175.97  & 12.86    & 0.16     & 1.76 & 0.73 \\
    MoTe$_2$ &  125.94  & 17.12    & 0.32     &  123.54  & 7.88     & 0.14     & 2.17 & 0.37 \\

    WS$_2$ &    244.18  & 29.96    & 0.28     &  240.99  & 19.91    & 0.18     &1.50 &  1.07 \\ 
    WSe$_2$ &   196.81  & 24.70    & 0.30     &  194.13  & 15.05    & 0.17     &1.64 &  0.83 \\
    WTe$_2$ &   137.32  & 18.71    & 0.32     &  135.27  & 9.30     & 0.15     &2.01 &  0.44 \\
  \end{tabular}
  \end{ruledtabular}
  \label{tab:xx-yy}
\end{table*}

As MX$_2$ monolayers are strained further~($\varepsilon > 4\%$), the stress-strain response deviates from the linear behavior. For large strains, the hexagonal symmetry is broken, with the stress developed upon loading in armchair direction~(Fig.~\ref{fig:stress}(a)) much larger than in the zigzag direction~(Fig.~\ref{fig:stress}(b)). Upon straining further, the stress continues to increase till it reaches a maximum, termed as the ultimate strength $\sigma^*$. Table~\ref{tab:xx-yy} presents the calculated values for the peak stress~$\sigma^*$ and the corresponding ultimate strain~$\varepsilon^*$ for all chemical compositions. It can be observed that in general,  the chalcogens of W~(WX$_2$) have larger moduli and tensile strength than that of Mo~(MoX$_2$). For the same transition metal, sulphides~(MS$_2$) are the strongest while tellurides~(MTe$_2$) are weakest. Moreover, Table~\ref{tab:xx-yy} also presents values for the anisotropy factor $\phi = \sigma^*_{\mbox{\scriptsize{AR}}} / \sigma^*_{\mbox{\scriptsize{ZZ}}}$, which measures the relative strength along the two high symmetry directions. We find that the anisotropy in stress response is inversely correlated with the strength of the monolayer sheets --- the MX$_2$ with lower Young's modulus and ultimate strength~(for example, tellurides) are characterized by larger anisotropy factors.

\begin{figure*}

  \subfigure[]
  {
    \includegraphics[scale=0.35]{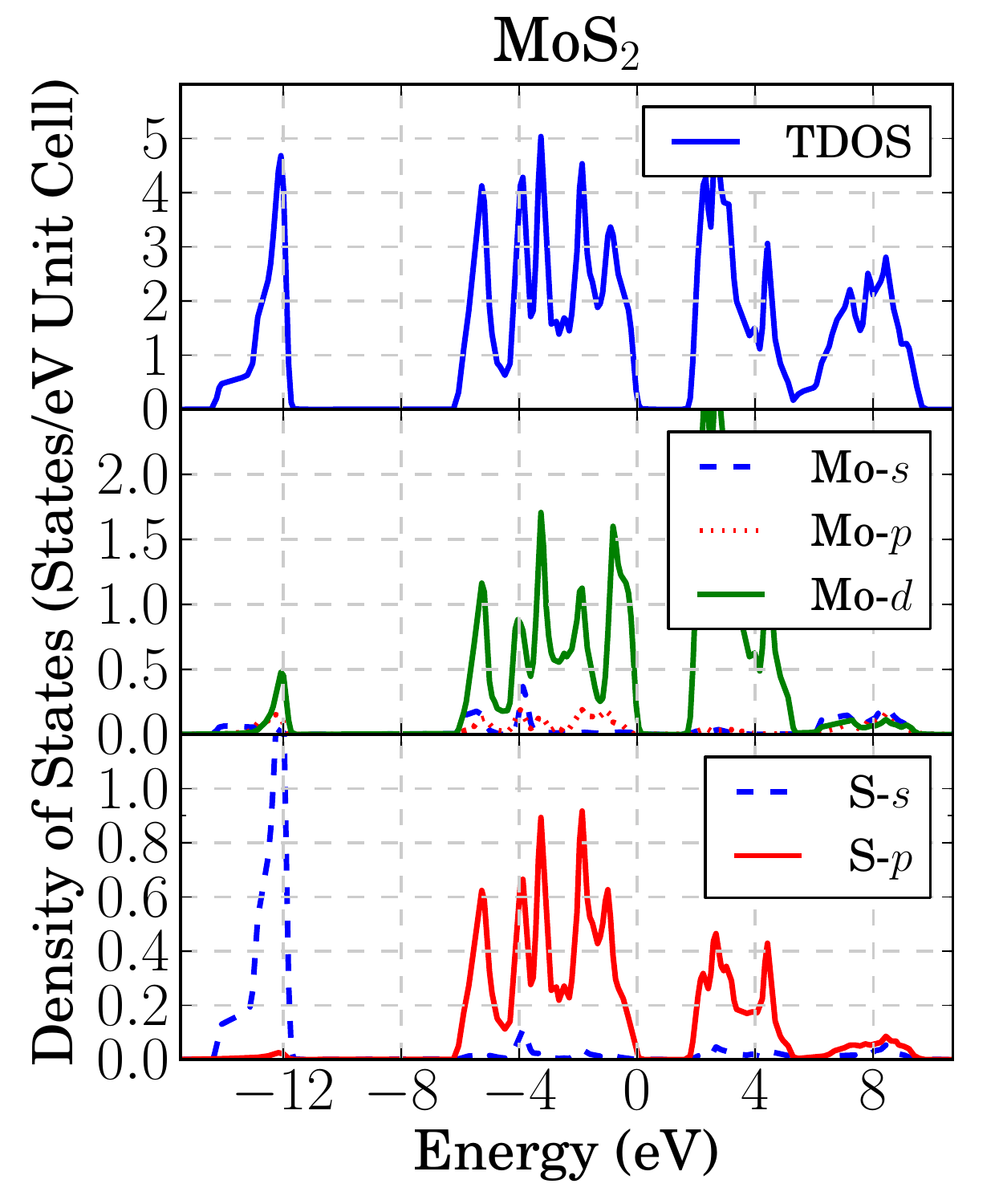}
    \label{fig:pdos_mos}
  }
  \subfigure[]
  {
    \includegraphics[scale=0.35]{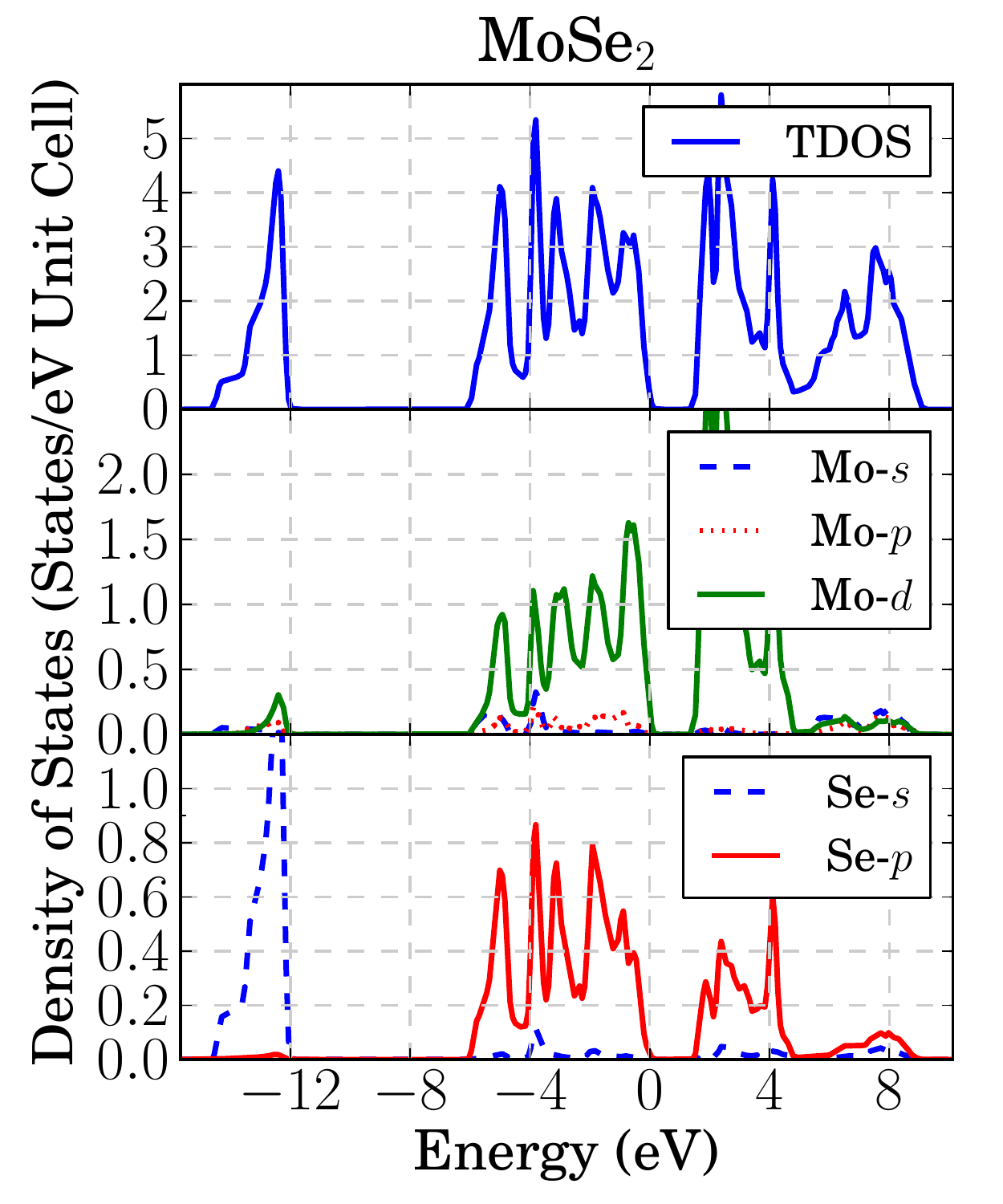}
    \label{fig:pdos_mose}
  }
  \subfigure[]
  {
    \includegraphics[scale=0.35]{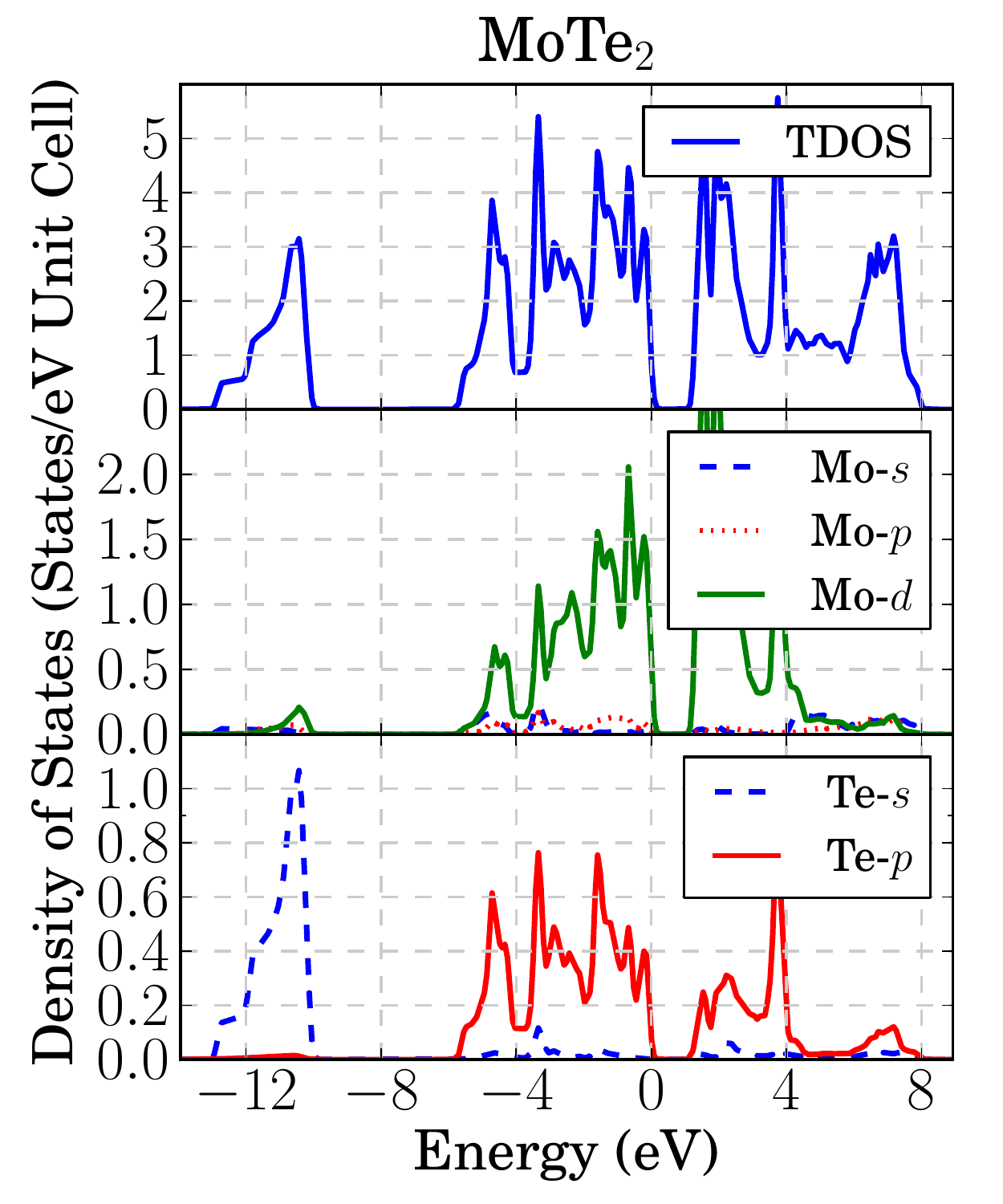}
    \label{fig:pdos_mote}
  }
  \subfigure[]
  {
    \includegraphics[scale=0.35]{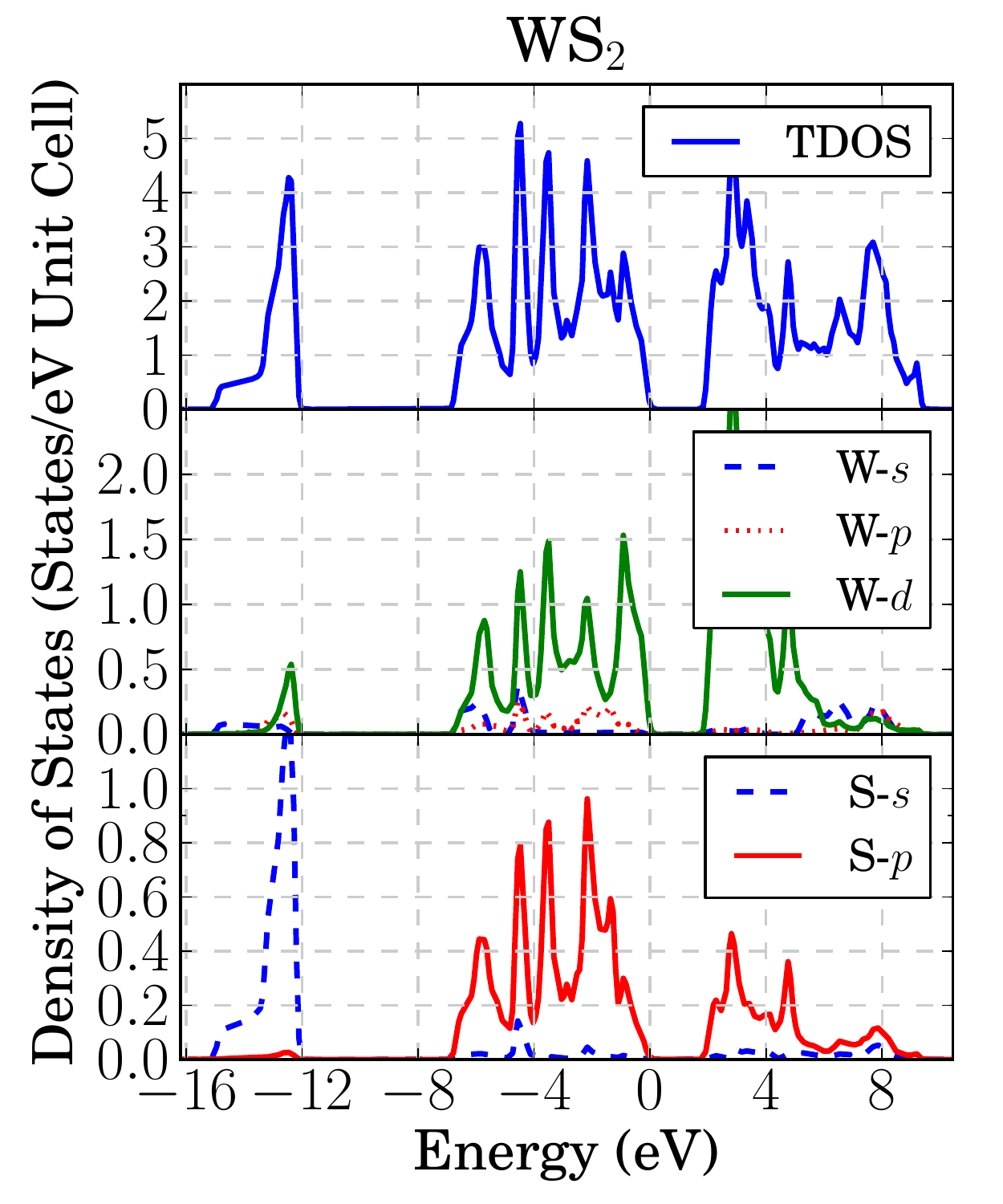}
    \label{fig:pdos_ws}
  }  \subfigure[]
  {
    \includegraphics[scale=0.35]{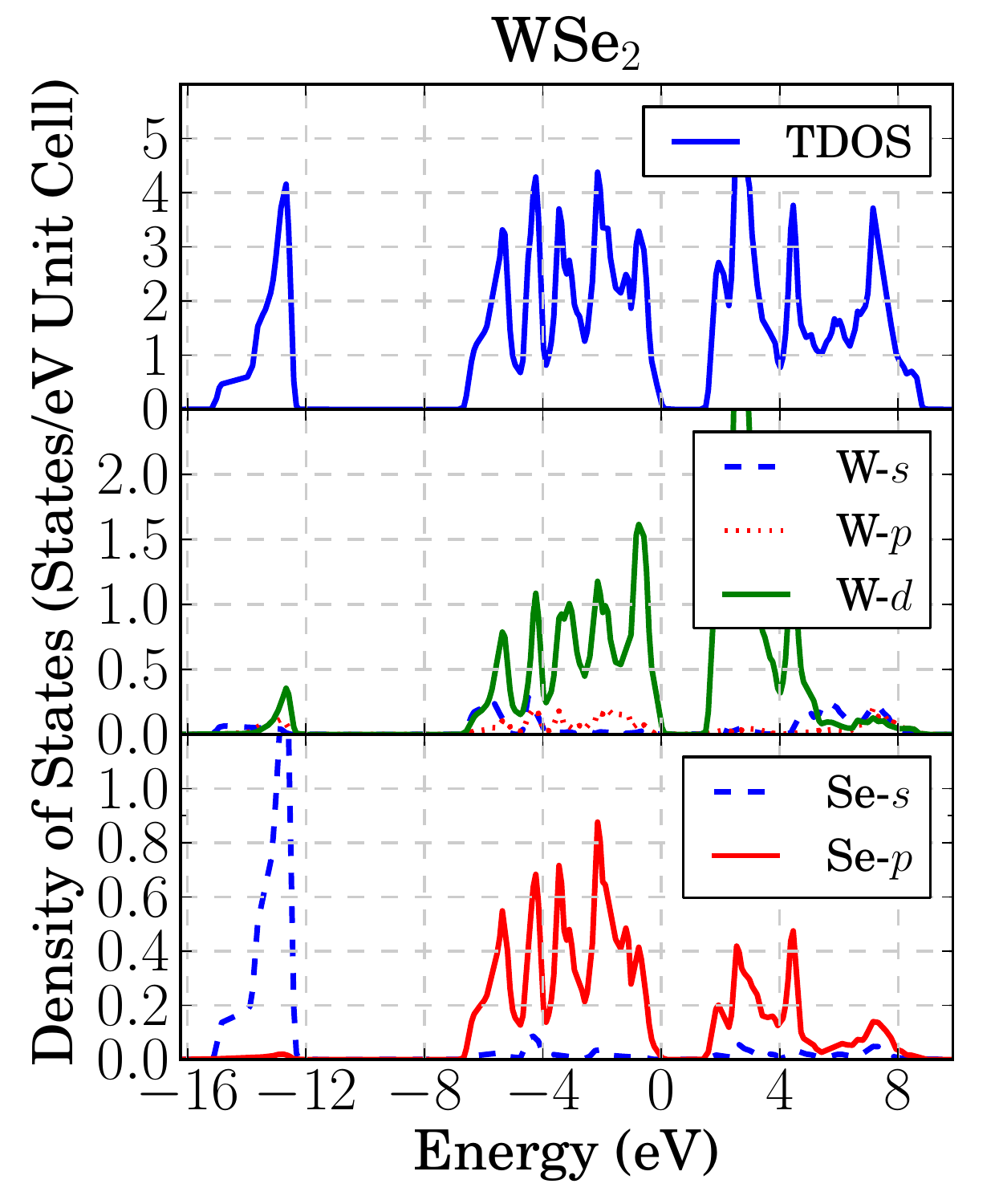}
    \label{fig:pdos_wse}
  }  \subfigure[]
  {
    \includegraphics[scale=0.35]{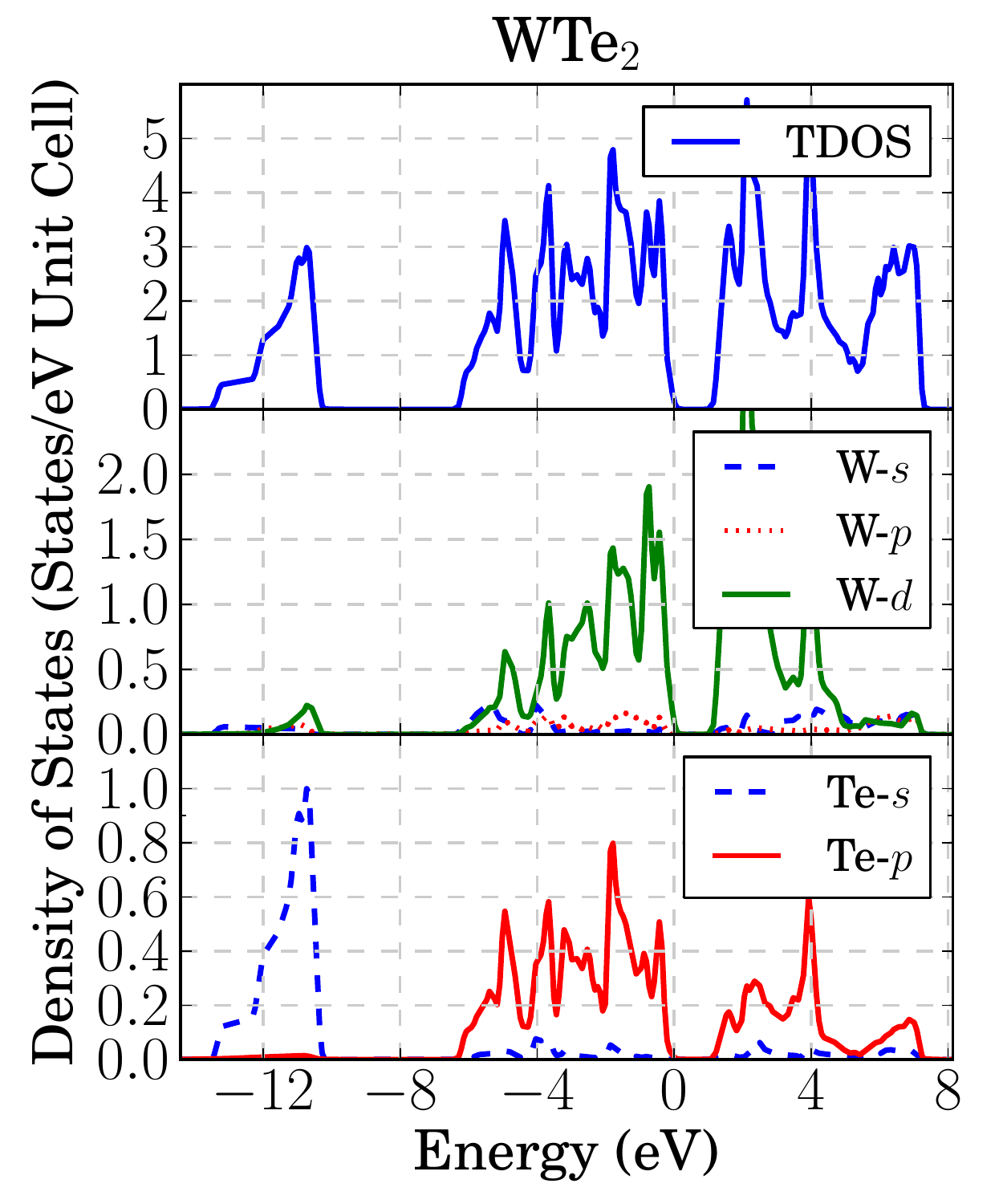}
    \label{fig:pdos_wte}
  }

  \caption{Total and projected density of states for strain-free monolayer MX$_2$~(M = Mo, W; X = S, Se, Te) TMDs. For the transition metal M and the chalcogen X, the density of states are projected onto $s$, $p$, $d$ and $s$, $p$ orbitals, respectively. The Fermi level is set to zero.}
  \label{fig:pdos}
\end{figure*}

Our calculated value for the average ultimate strength of monolayer MoS$_2$~(22 GPa) is in excellent agreement with the value of 23 GPa obtained by the atomic force microscopy measurements. Furthermore, the observed stress-strain response of the single-layer structure of the TMDs can be compared with their nanotube counterparts. For instance, density functional tight binding~(DFTB) simulations have predicted a Young's modulus of 209.7~GPa and 236.6~GPa for armchair and zigzag nanotubes of comparable diameter about 12.0~\AA, respectively, in close agreement with our results.\cite{Lorenz_jpcc_2012} Moreover, their ultimate strength was reported to be 29.1~GPa and 31.6~GPa, respectively, again in good qualitative agreement with our results. However the DFTB studies of MoS$_2$ nanotubes suggested similar value~($\sim$ 16\%) for the ultimate strain $\varepsilon^*$ corresponding to the ultimate strength for both armchair and zigzag directions, in contrast to our results where much bigger values of $\varepsilon^*$~($\geq$ 28\%) are observed for the armchair loading direction compared to $\varepsilon^*$~($\leq$ 19\%) for the zigzag loading direction~(see Table~\ref{tab:xx-yy}). 

It is also worth noting that a similar transition in stress-strain curve from an isotropic and linear stress response at small strains to anisotropic and nonlinear stress response at large strains has also been observed in the case of graphene.\cite{Liu_prb_2007} However, there is one remarkable difference: while TMDs and graphene both have hexagonal lattice, they exhibit a contrasting anisotropy in stress response. The peak stress in graphene for armchair tension is ~9\% lower  than that for zigzag direction, in contrast to the behavior seen in Fig.~\ref{fig:stress}. 

In order to identify the atomic-level origin of the mechanical behavior of various transition metal dichalcogenides considered here, we analyzed their electronic structures by using the Vienna \textit{ab initio} simulation package (VASP)\cite{Kresse_prb_1996} with LDA exchange correlation functional. Frozen-core projector-augmented wave (PAW) method was employed to describe the interaction between core electrons and valence electrons.\cite{Blochl_prb_1994} An energy cutoff of 400 eV was used for plane wave basis expansion.

The calculated total and projected density of states of unstrained MX$_2$ monolayer sheets are presented in Fig.~\ref{fig:pdos}.  Since the strength of the interatomic bonding and the consequent mechanical response of the materials to deformation are determined primarily by the occupied states just below Fermi level, we focus on the energy range 0 - 8 eV below the Fermi level. It can be seen that in this range, the outermost $d$ orbitals of the transition metal M overlap significantly with the outermost $p$ orbitals of the chalcogen X, indicating a strong $p$-$d$ hybridization. Furthermore, a careful examination of the peaks of these hybridized states yields a crucial insight on the composition dependence of the mechanical response seen in Fig.~\ref{fig:stress}. For instance, the peaks of $p$-$d$ orbitals in MoS$_2$ are located at -5.25, -4.02, -3.27, -1.93 and -0.83 eV below the Fermi level. In MoSe$_2$, these peaks shift toward the Fermi level by 0.28, 0.13, 0.15, 0.03 and 0.11 eV, respectively. The peaks shift further toward the Fermi level in MoTe$_2$ by 0.31, 0.52, 0.78, 0.30 and 0.08 eV, respectively. In addition to the change in peak positions, the height of peaks also changes as the composition is changed from MoS$_2$ to MoTe$_2$: the peaks of low-lying Mo-4$d$ states are lowered while the peaks of states close to Fermi level are enhanced. As seen in Fig.~\ref{fig:pdos}, the hybridized states for the chalcogens of W follow an essentially similar trend. 

The changes in the $p$-$d$ hybridized orbitals in monolayer TMDs can be quantitatively characterized by computing the center of $d$ bands below Fermi level, which is  defined  as
$$
E_d = \frac
{
  \int_{E_L}^{E_F} \mbox{PDOS}(E,d) \times E \, dE
}
{
  \int_{E_L}^{E_F} \mbox{PDOS}(E,d) \, dE
}
$$
where $\mbox{PDOS}(E,d)$ is the density of states projected onto the $d$ orbitals of transition metals Mo and W, the lower end energy $E_L$ is taken as -8 eV to cover the $p$-$d$ hybridized states and the Fermi level $E_F$ is set to zero. For MoS$_2$, MoSe$_2$ and MoTe$_2$, the calculated $d$ band centers $E_d$ are -2.68, -2.37, -1.94 eV, respectively. For WS$_2$, WSe$_2$ and WTe$_2$, $E_d$ is located at -2.98, -2.65, -2.18 eV, respectively. Since the hybridization of $p$-$d$ orbitals indicate a covalent electron sharing between M and X atoms, the deeper the center of the $d$ bands below the Fermi level, the stronger is the M-X bond. Consequently, the tellurides with shallow $E_d$ demonstrate lower Young's modulus and tensile strength compared to sulphides and selenides. Chalcogens of W have deeper $E_d$ than the chalcogens of Mo and are therefore stronger. The analysis of the locations of $d$ band centers presented here explains the observed composition dependence of the mechanical response of the monolayer TMDs as seen in Fig.~\ref{fig:stress}.

\begin{figure}  
  \includegraphics[scale=0.3]{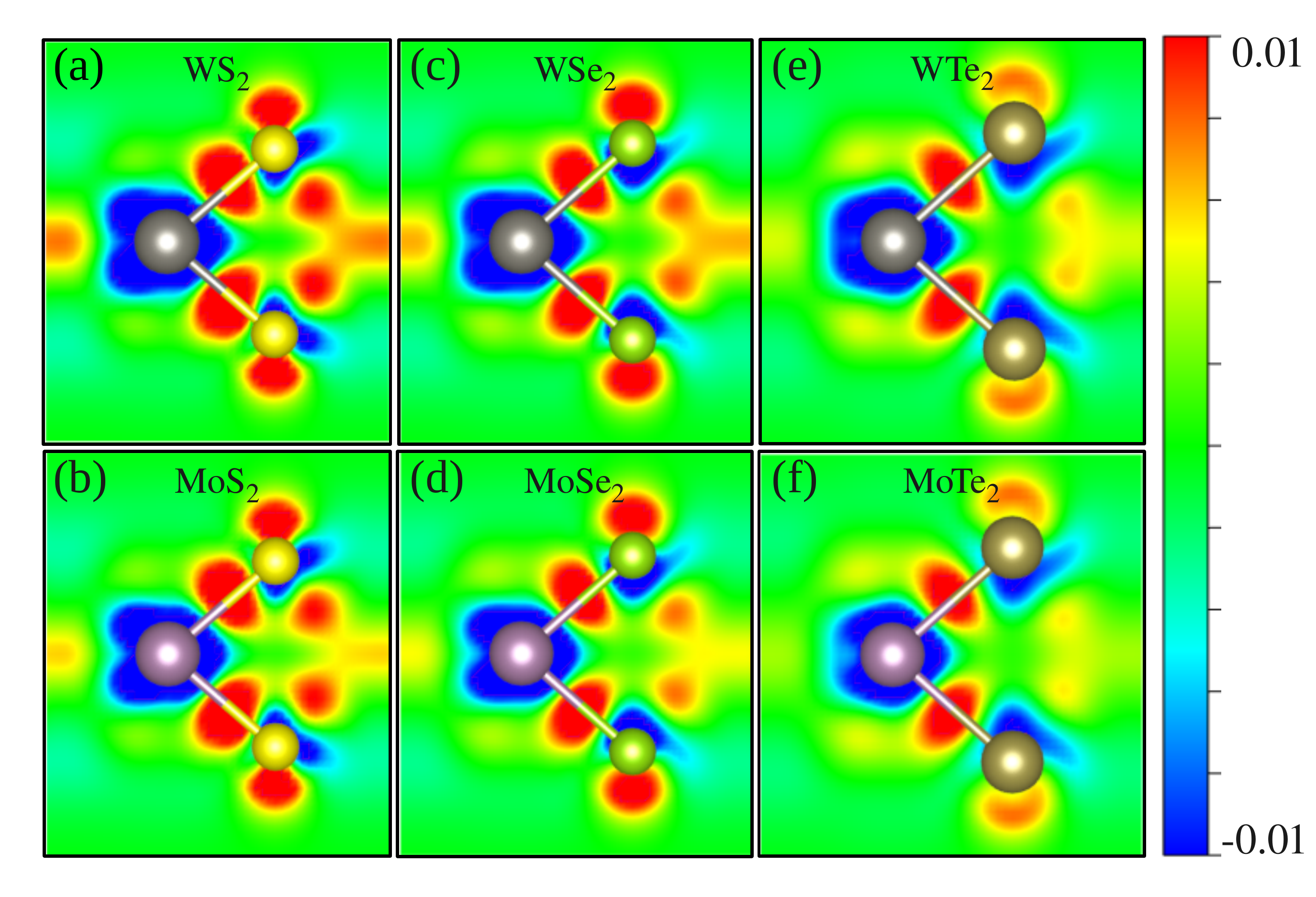}
  \caption{Bonding charge density for monolayer MX$_2$~(M = Mo, W; X = S, Se, Te) TMDs, obtained as the charge density difference between the valence charge density of the monolayer and the superposition of the valence charge density of the neutral constituent atoms. Red and blue colors indicate the electron accumulation and depletion, respectively. The color scale is in the units of $e$/Bohr$^3$.}
  \label{fig:charge}        
\end{figure}

\begin{figure}
  \subfigure[]
  {
    \includegraphics[scale=0.4]{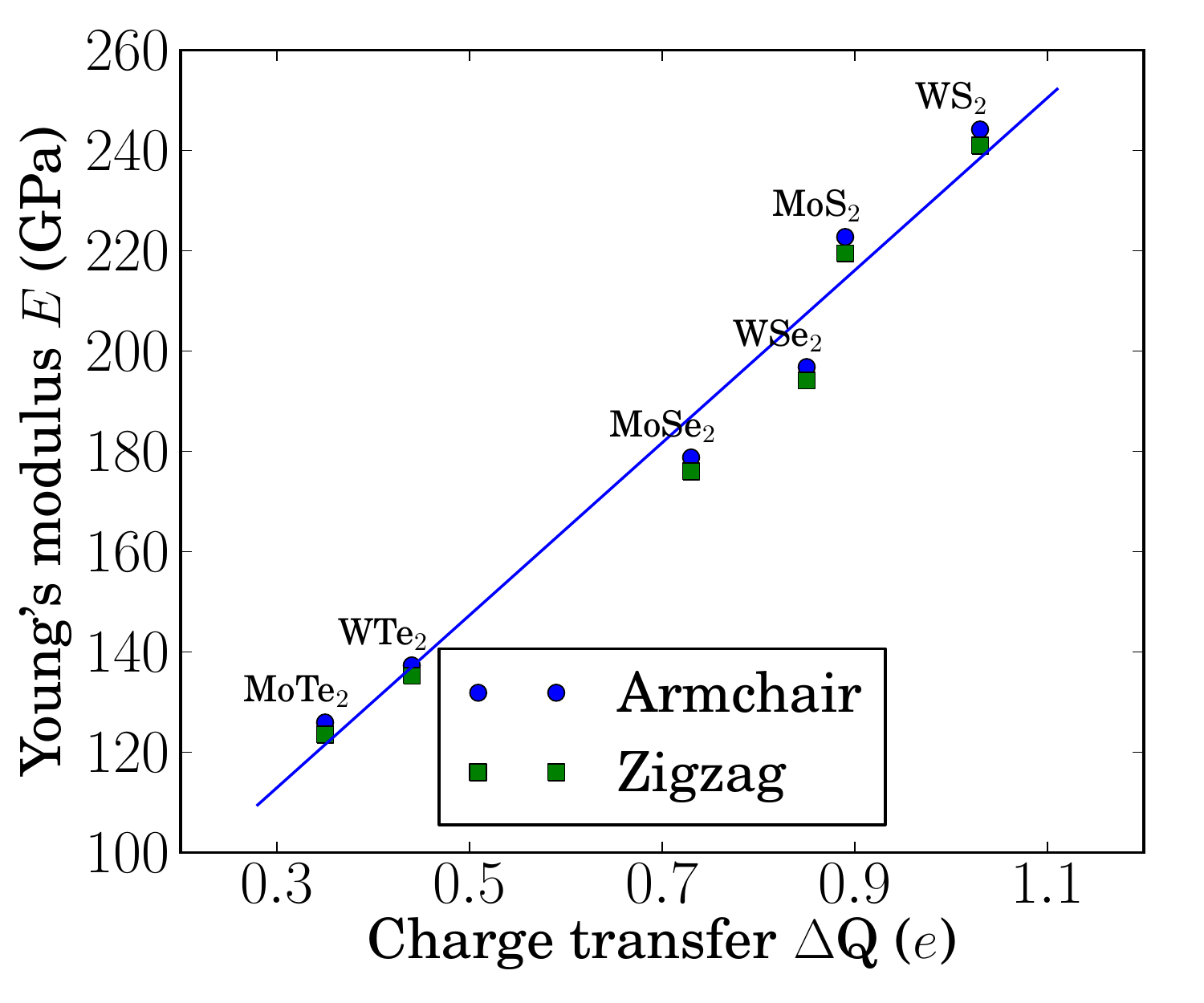}
    \label{fig:q_youngs}
  }
  \subfigure[]
  {
    \includegraphics[scale=0.4]{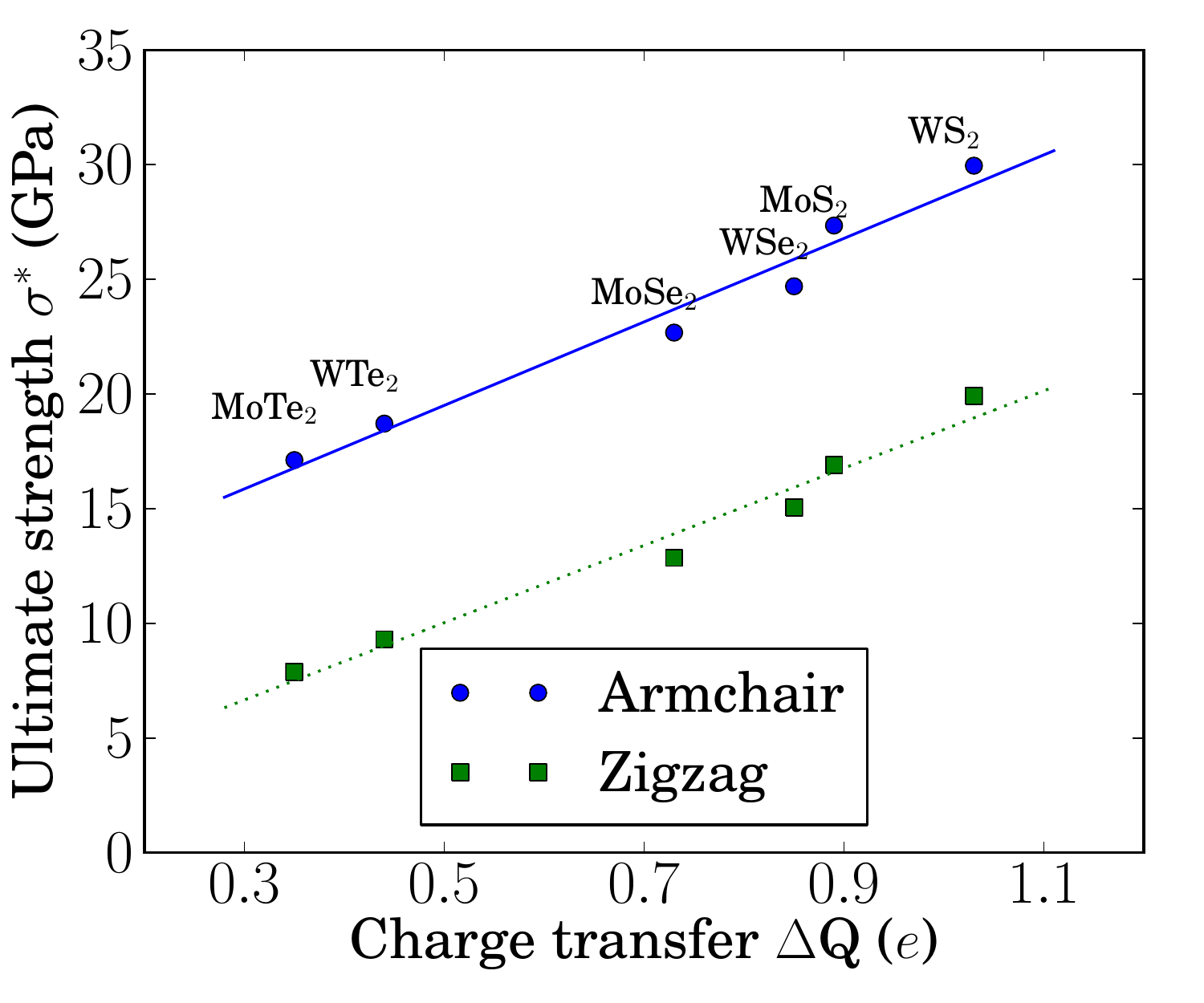}
    \label{fig:q_peak}
  }
  \caption{Variation of (a) the Young's modulus and (b) the ultimate strength of monolayer MX$_2$~(M = Mo, W; X = S, Se, Te) TMDs with the charge transfer $\Delta Q$ from transition metal M to chalcogens X.}
  \label{fig:transfer}
\end{figure}

In order to visualize the hybridized electronic states between the transition metals and chalcogens, we have further calculated the electronic charge distribution. Fig.~\ref{fig:charge} shows the bonding charge density in the plane passing through both the transition metal and chalcogen atoms. The bonding charge density is obtained as the difference between the valence charge density of strain-free MX$_2$ sheet and the superposition of the valence charge density of the constituent atoms.  A positive value~(red) indicates electron accumulation  while a negative value~(blue) denotes electron depletion. These bonding charge distributions clearly show the electron accumulation in the middle region of M and X. The amount of charge localized in this region qualitatively indicates the strength of the covalent M-X bond. It is evident that as the composition changes from sulphides to tellurides, the charge accumulation in the bonding regions gradually become less, indicating a weakening of M-X bond, lower Young's modulus and ultimate strength. 

\begin{table}
  \caption{Fitted values in the linear relation $\sim a + b \Delta Q$ to describe the dependence of the Young's modulus $E$ and ultimate strength $\sigma^*_{\mbox{\scriptsize{AR}}}$ and $\sigma^*_{\mbox{\scriptsize{ZZ}}}$ on the charge transfer $\Delta Q$.}
  \begin{ruledtabular}
  \begin{tabular}{ccc}
    & $a$ & $b$ \\
    \hline
           $E$ & 171.97 & 61.34 \\
           $\sigma^*_{\mbox{\scriptsize{AR}}}$  & 18.21 & 10.40 \\
           $\sigma^*_{\mbox{\scriptsize{ZZ}}}$ & 16.86 & 1.60 
  \end{tabular}
  \end{ruledtabular}
  \label{tab:transfer}
\end{table}

Fig.~\ref{fig:charge} also shows a large amount of charge transfer from the transition metal to the chalcogens.  This charge redistribution can be quantitatively estimated by computing the charge transfer from M to X. Table~\ref{tab:xx-yy} presents the magnitude of the charge transfer obtained using Bader charge analysis.\cite{Bader_book,Tang_jpcm_2009}  We depict the Young's modulus $E$ and ultimate strength $\sigma^*$ as a function of charge transfer $\Delta Q$ in Figs.~\ref{fig:q_youngs} and~\ref{fig:q_peak}, respectively. It is evident that the mechanical properties of transition metal dichalcogenides exhibit a linearly increasing relation with the charge transfer from transition metal atom to chalcogen atoms. By fitting the values of the Young's modulus and the ultimate strength with the charge transfer, this relationship can be simply described as $E(\sigma^*) \sim a + b \Delta Q$ where the values of fitted parameters $a$ and $b$ are listed in Table~\ref{tab:transfer}. These expressions provide an approximate and simple description of the mechanical properties of transition metal dichalcogenide monolayers sharing similar crystal structures.

\section{CONCLUSION}
In summary, we have investigated the relationship between the intrinsic mechanical response of the monolayer group VI TMDs to large elastic deformation, the atomic-level structure, and chemical composition. Our calculations demonstrate that the chemical composition of a monolayer TMD strongly influences its mechanical response to large elastic deformation, with the chalcogens of W exhibiting much greater ultimate strength than the chalcogens of Mo. The stress response of monolayer TMDs also depends on the crystal symmetry, with the armchair directions being consistently stronger across all chemical compositions. The origin of the observed mechanical behavior can be attributed to a strong hybridization between the outermost $p$ orbitals of the chalcogens and the $d$ orbitals of the transition metal. This hybridization leads to a redistribution of the electronic charge to the shared region between the transition metal and chalcogen atoms.  Our study clearly highlights the interplay between the atomic structure and the composition of monolayer TMDs that lends them a great strength to sustain large reversible deformations --- deformations large enough to modify their electronic structure as envisioned for novel optoelectronic and photonics applications. It should be emphasized that our first-principles investigation provides a realistic estimate of the intrinsic mechanical response of monolayer TMDs as measured by the nano-indentation experiments. In such experiments, the stressed region of the monolayer under the atomic force microscope tip can be expected to be defect-free and, therefore, the measured stresses can approach the ultimate strength. Therefore, our investigation of the stress-strain relation of monolayer TMDs presented here not only yields crucial insights into mechanical behavior of these materials but also can be compared with the experimental measurements.

\begin{acknowledgements}
J.L. and V.B.S. gratefully acknowledge the support of the Army Research Office
through Contract W911NF-11-1-0171. N.V.M. acknowledges the computational support from MASSIVE and NCI national facilities.
\end{acknowledgements}

\bibliography{ref.bib}

\end{document}